# Mechanisms of B Cell Synapse Formation Predicted by Stochastic Simulation

Philippos K. Tsourkas[*], Nicole Baumgarth[#], Scott I. Simon[*], Subhadip Raychaudhuri[*§]

[*]*Department of Biomedical Engineering*, [#]*Center for Comparative Medicine*
*University of California, Davis*
*One Shields Avenue, Davis, CA 95616, USA*

[§]Address correspondence to: raychaudhuri@ucdavis.edu

ABSTRACT

The clustering of B cell receptor (BCR) molecules and the formation of the protein segregation structure known as the "immunological synapse" appears to precede antigen (Ag) uptake by B cells. The mature B cell synapse is characterized by a central cluster of BCR/Ag molecular complexes surrounded by a ring of LFA-1/ICAM-1 complexes. Recent experimental evidence shows receptor clustering in B cells can occur via mechanical or signaling-driven processes. An alternative mechanism of diffusion and affinity-dependent binding has been proposed to explain synapse formation in the absence of signaling-driven processes. In this work, we investigated the biophysical mechanisms that drive immunological synapse formation in B cells across the physiological range of BCR affinity ($K_A \sim 10^6$-$10^{10}$ M$^{-1}$) through computational modeling. Our computational approach is based on stochastic simulation of diffusion and reaction events with a clearly defined mapping between our model's probabilistic parameters and their physical equivalents. We show that a diffusion-and-binding mechanism is sufficient to drive synapse formation only at low BCR affinity and for a relatively stiff B cell membrane that undergoes little deformation. We thus predict the need for alternative mechanisms: a difference in the mechanical properties of BCR/Ag and LFA-1/ICAM-1 bonds and/or signaling driven processes.

## INTRODUCTION

Specific recognition of foreign antigens by lymphocytes is central to the adaptive immune response. However, precisely how lymphocytes differentially respond to antigenic stimuli of varying type and strength remains unknown. Recent experimental evidence suggests antigen (or MHC-loaded peptides for the T cell-APC system) presentation on the restricted geometry of a two dimensional cell surface, together with recruitment of antigen into segregated clusters of receptor-ligand complexes, are one possible mechanism by which lymphocytes recognize and respond to antigen (1-9).

The clustering of receptor molecules on the cell surface and the formation of segregated protein structures increasingly is seen as an efficient mechanism of cellular information exchange during cell-cell interactions (10). In the immune system, such clustering and segregation of membrane-bound proteins is observed at the intercellular junction between lymphocytes and antigen presenting cells (APC) as they become adherent and engage in antigen recognition. Because of the resemblance to neurological synapses, these structures have been collectively termed "immunological synapses" (2-4).

The mature B cell synapse consists of a central cluster of B cell receptor/antigen (BCR/Ag) molecular complexes (sometimes also referred to as the central supramolecular activation cluster, or c-SMAC), surrounded by a ring of Lymphocyte function-associated antigen-1/Intercellular adhesion molecule-1 (LFA-1/ICAM-1) complexes (also known as the peripheral SMAC, or p-SMAC). This is the much studied canonical form of the immune synapse first observed at the inter-cellular junction between a T cell and an APC (1-4).

Although the mechanisms that drive synapse formation in T cells have extensively been modeled (11-18) since the pioneering work of Qi et al. (11), less is known about the mechanisms that govern synapse formation in B cells. Even though the structure of the B cell synapse is similar to that of the canonical form of the T cell synapse, factors such as receptor affinity, density and extracellular domain length of receptors vary significantly between T cells and B cells. It is thus thought the mechanisms that drive synapse formation in B cells may differ substantially from those in T cells (19).

B cell synapse formation is thought to be driven by signaling that results in changes in the membrane shape as a result of affinity-dependent BCR/Ag binding (8). An alternative proposal, the so-called diffusion-and-binding hypothesis, also has been proposed to explain B cell receptor clustering in signaling defective cells (19). According to this model, the synapse forms mainly by undirected diffusion of receptors into the contact zone, whereupon they bind with high affinity and stay in place or are eventually expelled by crowding at the center of the contact zone. This model has crucial limitations, however, in that (i) only a limited window of affinity ($K_A \sim 10^8$-$10^{10}$ M$^{-1}$) is used, whereas B cells can recognize antigen over a wider range of affinities ($K_A \sim 10^6$-$10^{10}$ M$^{-1}$), (ii) it does not consider the formation of the surrounding ring of LFA-1/ICAM-1 complexes.

In this study, we investigate the molecular mechanisms that drive B cell synapse formation by means of computational modeling. B cells are thought to respond to antigenic stimuli in a graded manner dependent on the affinity of BCR for the antigen encountered. We want to elucidate the mechanism of B cell synapse formation, which occurs over a wide range of affinities ($K_A \sim 10^6$-$10^{10}$ M$^{-1}$), by systematically investigating the effect of biophysical parameters such as bond stiffness, bond length, diffusion, and membrane rigidity on synapse formation.

Our approach consists of a stochastic, agent-based, computer model of a B cell-APC interaction in which individual molecular events such as diffusion and reaction are simulated using probabilistic rules. Individual parameter values of affinity and molecular concentrations can be varied in a controlled manner in successive *in silico* experiments to identify their contribution to the process of synapse formation. Our model is based on available experimental data and allows us to gauge the effects of changing individual parameters to predict B cell response in a rapid and computationally efficient manner.

Our results show that the basic diffusion-and-binding mechanism is indeed sufficient to explain the clustering of BCR/Ag complexes and LFA-1/ICAM-1 ring formation, but only if the affinity of BCR for antigen is less than that of LFA-1 for ICAM-1 and no membrane deformation takes place (as is the case at the onset of synapse formation). For high affinity BCR/Ag binding, our model predicts that it is necessary for BCR/Ag bonds to be stiffer than LFA-1/ICAM-1 bonds for a synapse to form. However, when significant membrane shape change is allowed in our model, an additional mechanism, either in the form of a difference in extracellular domain length between BCR/Ag and LFA-1/ICAM-1 complexes, or a signaling driven process, becomes necessary for synapse formation. One such signaling-driven process, is a shift in the affinity of LFA-1 from an initial low to a high affinity state as a result of BCR signaling (6,7,9).

The organization of the paper is as follows: The specific features of our Monte Carlo technique and the details of the simulation procedure are described in the model section. The results section is divided into two parts – (i) no membrane deformation and (ii) membrane deformation is considered. The mechanisms of synapse formation are addressed as the affinity is varied over the physiological range. The significance of antigen concentration and diffusion on synapse formation is also reported. Finally, the main findings of this work and potential refinements to our model are summarized in the discussion section.

# MODEL

Background

A Monte Carlo method was applied to model B cell synapse formation. In these methods, the molecular population is randomly sampled to undergo events such as diffusion and reaction, with its status updated at every time step. Monte Carlo methods have been successfully employed in the past to understand immune cell receptor-ligand binding, clustering and signaling (13, 18, 20-24). Our model's distinguishing features are: (i) the use of probabilistic rate constants in our model instead of an energy-based Metropolis algorithm (ii) explicit spatial simulation of molecules to allow the modeling of spatial crowding and exclusion, and (iii) treatment of diffusion of receptor-ligand complexes. The explicit spatial simulation of molecules allows the modeling of crowding and exclusion effects that are potentially important in synapse formation but cannot easily be captured by differential equation-based models, particularly if more than one molecular species is present. The discrete nature of our model also eliminates the need to make assumptions about the continuity of molecular concentrations, which may not be valid at low antigen concentration. This is particularly relevant in light of the fact that antigen concentration is usually low at the onset of an immune response.

Model setup

As shown in Figure 1, the section of the B cell – APC system we wish to model is the region of closest approach between the two cells, where the distance between the membranes is small enough to allow binding between molecules on opposite surfaces. The cell membranes are modeled as two Cartesian lattices, each discretized in an $NXN$ grid of nodes. We assume the membranes initially have a spherical curvature, which, in the absence of external forces, both cells would tend towards in order to minimize surface energy. The total vertical separation distance $z$ between the two surfaces at any given point $(x, y)$ is given by $z=z_1+z_2$, as in Figure 1, with the half-heights $z_1$ and $z_2$ each given by:

$$z_i(x,y) = \frac{z_o}{2} + R_i - \left(R_i^2 - ((x-x_0)^2 + (y-y_0)^2)\right)^{1/2} \qquad (1)$$

At the center of the contact zone ($x=x_o$, $y=y_o$), the vertical separation between the two surfaces is at its minimum value, $z=z_o$. We also can simulate a cell-bilayer system such as the one used in many synapse experiments (6-8) in the limit as $R_2 \to \infty$ and $z_2(x,y) \to z_o/2$.

The size of the region we have chosen to simulate in our model is 3 μm square, which is large enough to include the entire region over which binding can occur (see Fig. 2), and also larger than typical experimentally observed synapse diameters of ~2 μm (6,7). In addition, this area is believed to be sufficiently large such that a zero net flux condition exists at the boundaries, which in our agent-base simulator is simulated by means of fully reflecting boundaries. Only one molecule can occupy a node in our simulation, so we choose a nodal spacing equal to a membrane protein molecule's exclusion radius, ~10 nm (resulting in 300X300 nodes). The various parameters that relate to the spatial dimensions of our model are listed in Table 1.

Simulation procedure: reaction and diffusion "moves"

At the start of a simulation run, molecules are uniformly distributed over the two surfaces at random. The molecular species represented are BCR and LFA-1 on the B cell surface and antigen and ICAM-1 on the APC or bilayer surface. At every time step in the simulation, molecules from the population are individually sampled at random to undergo either diffusion or reaction events. Which of the two events a selected molecule will undergo is determined by means of a coin toss with probability $p_{d/r}$, which is set to 0.5.

Reaction

If a molecule has been selected to undergo a reaction, the first step is to check for a complementary molecule (since only two reactions are possible: BCR+Ag↔BCR/Ag and LFA-1+ICAM-1↔LFA-1/ICAM-1) at the same node on the opposite surface. If that is the case, a random number trial with probability $p_{on(i)}$ is performed to determine if the two molecules will bind together and form a molecular complex. The binding probability is specific to the reaction and the subscript $i$ refers to the BCR/Ag reaction when $i=BA$ and the LFA-1/ICAM-1 reaction when $i=LI$. We assume the probability of bond formation depends on the intermembrane distance $z$ in accordance to the well-known linear spring model (25,26), and by replacing the rate constant $k_{on}$ with the probability $p_{on}$, we obtain the following probability density function:

$$p_{on(i)}(z) = p_{on(i)}^{max} \exp\left(-\frac{\kappa_i'(z-z_{eq(i)})^2}{2k_B T}\right) \quad (2)$$

The bond is modeled as a mechanical spring with stiffness $\kappa'$ and equilibrium length $z_{eq}$, while $k_B$ denotes the Boltzmann constant ($1.34*10^{-23}$ J/K) and $T$ the temperature (~300 K). The probability of binding is greatest at the point $z(x,y)=z_{eq}$, which will be the center of contact zone when $z_{eq}=z_o$, as in Figure 2A. In formulating our model, we assumed that as the two cells move closer to one another and $z_o$ decreases, the first binding event is likely to occur when $z_o$ approaches the value of $z_{eq}$ of one of the species, after which the cells stop moving towards each other. In our simulations we thus set $z_o$ equal to either $z_{eq(BA)}$ or $z_{eq(LI)}$, depending on the circumstances.

In the case where the molecule selected to undergo a reaction is a two-molecule complex, however, the result is the possible dissociation of the complex into its component molecules. The dissociation probability $p_{off(i)}$ is given by:

$$p_{off(i)}(z) = p_{off(i)}^{min} \exp\left(\frac{(\kappa_i - \kappa_i')(z-z_{eq(i)})^2}{2k_B T}\right) \quad (3)$$

Without loss of generality, we can set $\kappa_{(i)}=2\kappa'_{(i)}$ so that the exponential in Eq. 3 is the same as that in Eq. 2 but with a positive sign in front. In contrast to $p_{on}$, $p_{off}$ is a minimum at $z=z_{eq}$, increasing away from this point, as in Figure 2B, where we also see that $p_{off}$ cannot exceed 1.0.

Since $p_{on}$ and $p_{off}$ are analogous to $k_{on}$ and $k_{off}$, we can obtain the probabilistic analog to the association constant $K_A$, denoted as $P_A$, by dividing Eq. 2 by Eq. 3 and setting $\kappa_{(i)}=2\kappa'_{(i)}$:

$$P_{A(i)}(z) = \frac{p_{on(i)}^{max}}{p_{off(i)}^{min}} \exp\left(-\frac{(\kappa_i(z-z_{eq(i)})^2}{2k_B T}\right) = P_{A(i)}^{max} \exp\left(-\frac{\kappa_i(z-z_{eq(i)})^2}{2k_B T}\right) \quad (4)$$

Affinity is simulated by the ratio $P_{A(i)}^{max} = p_{on(i)}^{max}/p_{off(i)}^{min}$ which plays much the same role as $K_A$ in governing equilibrium binding in our simulation. Individually varying $p_{on}^{max}$ and $p_{off}^{min}$ while keeping the ratio $P_A^{max}$ constant changes the time scale of the simulation but not the equilibrium behavior. A typical plot of $P_A$ is shown in Figure 2C, and it is important to note that $P_A$ is *not* a probability but a probability *ratio* we use to model affinity. The mapping between $K_A$, $k_{on}$, $k_{off}$ and $P_A^{max}$, $p_{on}^{max}$, $p_{off}^{min}$, respectively, is given in the Appendix.

Diffusion

On the other hand, if a molecule has been selected to undergo diffusion, a random number trial with probability $p_{diff(i)}$ is used to determine if the diffusion move will occur successfully. If the trial is successful, the selected molecule will "hop" to any neighboring node with equal probability (there are up to four possibilities, as we simulate the cell surface as a 2-D square lattice). Because only one molecule is allowed to occupy a node at any given time, the diffusion hop will only occur if the selected neighboring node is unoccupied (or both nodes in the case of a complex). The mapping between $p_{diff(i)}$ and the diffusion coefficient $D$ is given in the Appendix.

Membrane free energy and deformation

Our simulation also allows the modeling of changes in the originally spherical membrane shape as a result of receptor-ligand binding. We use the membrane free energy used by (11) and (18), which has the following form:

$$E = \sum_{i=1}^{2} \frac{\kappa_i}{2} \iint C_i (z - z_{eq(i)})^2 \, dxdy + \frac{1}{2} \iint \left[ \gamma (\nabla z)^2 + \beta (\nabla^2 z)^2 \right] dxdy \tag{5}$$

The first term in the equation relates to the energy associated with binding, which is a function of the concentration of $C_i$ of BCR/Ag and LFA-1/ICAM-1, while the other two terms relate to the energy associated with membrane tension ($\gamma$) and bending rigidity ($\beta$), respectively. The change in the membrane separation distance $z$ is modeled according to the well-known Landau-Ginzburg formulation in the manner of Qi et al. (11), which for the geometry used here has the form:

$$\frac{\partial z}{\partial t} = M \left( -\sum_{i=1}^{2} \kappa_i C_i (z - z_{eq(i)}) + \gamma \nabla^2 z - \beta \nabla^4 z \right) \tag{6}$$

The constant $M$ relates the time scale of membrane deformation relative to that of receptor-ligand binding, such that for small $M$, the membrane will essentially retain its shape for the duration of the simulation. Because the length scale of membrane deformation is considerably larger than that of a protein's exclusion radius (~100 nm instead of ~10 nm), for the purpose of calculating $z$ we coarse-grain the $NXN$ membrane surface lattice into 10 node X 10 node subdomains over which $z$ is constant. The concentration of complexes in each of these subdomains is then calculated and entered in the discrete form of Eq. 6.

Monte Carlo time step

In our algorithm, a number $S$ of diffusion/reaction trials is performed during every time step, at the end of which the membrane height is adjusted using Eq. 6 with Dirichlet boundary conditions in accordance with Fig. 1. The number of coin tosses and moves $S$ is set equal to the total number of molecules (free and complexes) present in the system at the beginning of each time step, and the simulation is run for a number of time steps $T$. A summary of our Monte Carlo algorithm is shown in Figure 3.

Model parameters

Our investigation strategy consists of successive virtual experiments in which individual parameter values are incrementally varied to determine the role each parameter plays in synapse formation. The full list of parameters used in our model is given in Table 2. This list includes all biological parameters whose values can be varied in our simulations, but does not include the spatial parameters listed in Table 1.

Many of the model parameters in Table 2 do not appear to vary significantly during physical experiments, however, or have values that can be found in the literature. Experimentally measured parameter values available in the literature that are relevant to our model of synapse formation are listed in the two columns on the left hand side of Table 3. In some cases, it is necessary to map the experimental value into the probabilistic analogs used by our model (see Appendix), as is the case with $K_A$, $k_{on}$, $k_{off}$ and $P_A^{max}$, $p_{on}^{max}$, $p_{off}^{min}$, respectively. The adapted forms of the experimental parameter values found in the literature which we use in our simulations are listed in the two columns on the right hand side of Table 3.

In the case of the diffusion coefficients of free molecules, these are found in the experimental literature to be ~0.01 $\mu m^2$/sec (27), with little variation between species. We thus collectively group the individual diffusion probabilities of the free molecule species from Table 2 ($p_{diff(B)}$, $p_{diff(A)}$, $p_{diff(L)}$, $p_{diff(I)}$) into a single parameter $p_{diff(F)}$, and likewise group the individual diffusion probabilities of the complexes, $p_{diff(BA)}$ and $p_{diff(LI)}$, into a single parameter $p_{diff(C)}$. A diffusion coefficient of the order of 0.01 $\mu m^2$/sec approximately corresponds to $p_{diff(F)}=1.0$ (see Appendix), while $p_{diff(C)}$ is assumed to be unknown and therefore variable.

In addition to using available literature values wherever possible, the number of variable parameters can be further reduced by making certain appropriate assumptions regarding the number of free BCR and LFA-1 molecules initially present on the B cell surface, $B_0$ and $L_0$. For instance, assuming a typical membrane protein molecule distribution of $10^5$ molecules/cell (12,29), and using a typical lymphocyte radius of 6 μm (resultant area ~ 450 $\mu m^2$), the average molecular density is ~ 220 molecules/$\mu m^2$. For a contact area of 9 $\mu m^2$, this means $B_0=L_0=2000$.

It also is possible to estimate the equilibrium length of the BCR/Ag complex, even though an exact number is not available in the literature. In the *in vitro* experiments we are basing our model on, the antigen molecules are part of antigen-antibody immune complexes loaded onto Fc receptors (5-7), which would indicate a minimum extracellular length comparable to that of LFA-1/ICAM-1 complexes, i.e. ~42 nm. However, it also is possible that in certain *in vivo* situations the antigens on the APC surface are fragments less than 1 nm in length, which would set the lower bound on the length of BCR/Ag complexes to the typical length of an antibody molecule, 22-23 nm (30,31). In our investigation we thus perform experiments where the length of the BCR/Ag complexes is set to either 22 or 42 nm.

When these assumptions and simplifications have been entered into our model, the list of variable parameters reduces to that shown in Table 4. These are the parameters for which unique literature values have not been found, such as $\kappa_{BA}$ and $z_{eq(BA)}$, (and which may well vary), or those that are varied in actual synapse experiments, such as antigen molecule number $A_0$ and BCR affinity $P_{A(BA)}^{max}=p_{on(BA)}^{max}/p_{off(BA)}^{min}$. These also are therefore the parameters we focused on as possible driving factors of B cell synapse formation.

# RESULTS

I. Insignificant Membrane Deformation

*Affinity difference between BCR and LFA-1 can drive synapse formation at low BCR affinity*

In nature, BCR affinity for antigen is critical in determining B cell response (32-36). In our simulations, we found that affinity can be a leading driver of synapse formation. In Figure 4, the affinity of BCR for antigen is varied across four orders of magnitude, from $K_A=10^5$-$10^8$ M$^{-1}$, while the affinity of LFA-1 is fixed at $K_A=10^7$ M$^{-1}$. In Figure 4*A*, the affinity of BCR is clearly too low for a synapse to form, even though the traces of one are discernible. In Figure 4*B*, however, we can clearly see the difference in affinity between BCR/Ag and LFA-1/ICAM-1 is sufficient to produce patterns similar to experimentally observed B cell synapses (6-8). No such pattern is observed when the affinities are equal (Fig. 4*C*), while an inverted pattern forms when BCR affinity exceeds LFA-1 affinity (Fig. 4*D*).

Our explanation for this behavior is as follows: Initially, the various molecules, all in the free state, are scattered uniformly over the cell surfaces. Because the region where binding is possible (defined by $p_{on}>0$ and $p_{off}<1$ in Fig. 2) is relatively small compared to the overall region of contact, at the start of our simulations most molecules are located outside the region of binding. Those molecules that happen to be initially located in this region, however rapidly bind and form complexes which tend to stay in place as the likelihood of rebinding upon dissociation in this region is high (this explains the presence of some LFA-1/ICAM-1 complexes in the center of the synapse). The synapse pattern forms as free molecules from the periphery randomly drift into the zone of binding until they eventually find a binding partner and form a complex.

If the complexes have relatively low diffusivity, as is the case in Figure 4, a ring-like pattern results as the complexes tend to stay near where they formed, at the edge of the region of binding. Over time these complexes may break up, and some of the newly freed molecules are equally likely to drift further into the zone of binding. As the probability of binding is higher and that of dissociation lower in the interior of the contact zone due to the curvature of the membrane, the ring-like pattern becomes more cluster-like over time. In Figure 4*B*, BCR has a lower affinity and higher $k_{off}$ than LFA-1, so that it forms a cluster at a faster rate than LFA-1 and thereby producing a synapse. The situation is reversed in Figure 4*D*, while a purely random pattern is produced in Figure 4*C* as the affinities and off-rates are equal. This is exactly the diffusion-and-binding mechanism with the addition that synapse formation is mainly driven by the difference in affinity (and in particular $k_{off}$) and the key provision that complexes have a lower diffusivity than free molecules. The details of this mechanism, together with the time evolution of the patterns in Figure 4 are discussed in the Supplemental Materials section.

A similar picture emerges when the BCR/Ag complex length is set to its theoretical minimum value, $z_{eq(BA)}=22$ nm, the main difference being that the pattern produced with $K_A=10^7$ M$^{-1}$ (Fig. 4*C*) is somewhat less purely random as the BCR/Ag complexes are distributed entirely within the ring of LFA-1/ICAM-1 complexes (although the distribution is still sufficiently random so that we do not discern a proper synapse pattern). Above this affinity value, the BCR/Ag complexes form a ring as in Fig. 4*D*. However, because of their greater length, the LFA-1/ICAM-1 complexes are excluded from the center so that the ring of

BCR/Ag complexes is located entirely within the ring of LFA-1/ICAM-1 complexes, producing a pattern of two concentric rings instead of the inverted synapse in Fig. 4*D*. From these results, it appears that differences in affinity alone are sufficient to form a synapse when BCR affinity is lower than LFA-1 affinity; however it also is clear that an additional mechanism is necessary to produce synapses at higher values of BCR affinity.

*BCR/Ag bond stiffness is crucial to synapse formation at high BCR affinity*

Given there are several antibody molecules that serve as B cell receptors for antigen, and that these receptors encounter a wide variety of antigens, it is reasonable to assume the stiffness of any given BCR/Ag bond ($\kappa_{BA}$) will vary in addition to the length and affinity. This notion is supported by the fact we did not find a value for the bond stiffness of the BCR/Ag complex in the literature, in contrast to LFA-1/ICAM-1. In our simulations we have discovered that differences in bond stiffness between BCR/Ag and LFA-1/ICAM-1 complexes play a key role in aiding synapse formation, especially at high BCR affinity.

First, we found that increasing the stiffness of the BCR/Ag bond above that of the LFA-1/ICAM-1 bond can result in synapses forming over the entire range of simulated BCR affinity values from ($K_A=10^6$ M$^{-1}$-$10^{10}$ M$^{-1}$). For example, our simulations show that a synapse such as the one in Figure 4*B* will form at a BCR affinity value of $K_A=10^7$ M$^{-1}$ when the stiffness of the BCR/Ag bond is set to $\kappa_{BA}=160$ μN/m (with $\kappa_{LI}=40$ μN/m), where we previously obtained a purely random pattern when the bonds had equal stiffness (Fig. 4*C*). At higher BCR affinity values, a greater increase in BCR/Ag bond stiffness is necessary to produce a synapse, as can be seen in Figure 5. Our explanation for this mechanism of synapse formation is that increasing bond stiffness narrows the width of the peak in Figure 2 (without lowering it), thereby reducing the radius of the zone of binding for BCR and forcing the BCR/Ag complexes into a smaller area. BCR/Ag complexes thus form closer to the center than LFA-1/ICAM-1 complexes, resulting in a concentric pattern with BCR/Ag complexes on the inside and LFA-1/ICAM-1 complexes on the outside. If the bond stiffness is sufficiently high for a particular BCR affinity value, the ring of BCR/Ag complexes will compress into a cluster, resulting in a synapse. With increasing BCR affinity, it takes increasingly longer for the BCR/Ag complexes to collect into a cluster (since $p_{\text{off(BA)}}^{\min}$ decreases), and thus a stiffer bond is needed to produce a synapse. The minimum BCR/Ag bond stiffness needed for synapse formation for the full range of BCR affinity values is given in Table 5 for both $z_{(eq)BA}=42$ nm and $z_{(eq)BA}=22$ nm and is also plotted in Figure 5.

*A threshold number of antigen molecules is needed for synapse formation*

The number of antigen molecules initially expressed on the APC membrane ($A_0$), together with affinity, are perhaps the most tightly regulated immunological parameters (6,7). The effect of varying the initial number of antigen molecules on synapse formation is shown in Figure 6. As depicted in Figure 6*A-C*, there was not significant change in the synapse pattern as the initial number of antigen molecules was decreased from 2000 to 500, except in the density of BCR/Ag complexes in the cluster. However, synapses did not form when the number of antigen molecules was reduced below 500, as can be seen in Figure 6*D*.

This situation persists at all BCR affinity values, with no synapses observed to form below $A_0=500$. Our model thus confirms the existence of a threshold number of antigen molecules needed for a synapse to form, and that this threshold is furthermore independent of affinity. Increasing the number of antigen molecules beyond the threshold, however, has only a relatively minor effect on the frequency of synapse formation and synapse quality.

*A shift in the affinity of LFA-1 as an alternative mechanism of synapse formation*

It has recently been hypothesized that LFA-1 on the B cell surface may initially be in a low affinity state prior to contact with the APC, and that it changes conformation into a high affinity state after outside-in signaling following BCR activation upon antigen ligation (6,7,9). As part of our investigation, we explored whether such a mechanism might promote the process of synapse formation. To model the process, we initially set the affinity of LFA-1 to a low value of $K_A \approx 10^4$ M$^{-1}$, and switch it to the high-affinity value of $K_A = 10^7$ M$^{-1}$ after either a critical time is reached or a critical number of BCR/Ag complexes have formed. The synapses produced with the shift in LFA-1 affinity did not differ in any noticeable way from synapses produced without it, however, nor was there any difference in the minimum BCR/Ag bond stiffness or antigen concentration needed to produce a synapse.

*Low complex diffusivity is crucial to synapse formation*

Our investigation further revealed that diffusion plays a central role in synapse formation. As mentioned earlier, we varied the probability of diffusion of molecular complexes, $p_{\text{diff}(C)}$, relative to the diffusion coefficient of free molecules, $p_{\text{diff}(F)}$, which was fixed at 1.0 (see Table 3). Our results revealed differences in diffusivity between free molecules and molecular complexes are crucial to the formation and stability of the synapses formed by the various mechanisms previously mentioned.

Specifically, our results indicate that the probability of diffusion of complexes must be at least two orders of magnitude lower than that of the free molecules for anything resembling a proper synapse to form, irrespective of other parameter values. This is clearly shown in Figure 7, where the ordered structure formed in Figure 7*A*, where $p_{\text{diff}(C)}$=0.01, has significantly deteriorated in Figure 7*B*, where $p_{\text{diff}(C)}$=0.1, and still further in Figure 7*C*, where $p_{\text{diff}(C)}$=1. This deterioration in the synapse pattern with increasing complex diffusivity is observed across the entire range of BCR affinity values and antigen molecule numbers, regardless of the synapse forming mechanism used (i.e. simple diffusion-and-capture, LFA-1-affinity-shift dependent, bond property dependent). We also note the number of BCR/Ag complexes dropped significantly as we increased $p_{\text{diff}(C)}$, from ~150 in Figure 7*A* to ~50 in Figure 7*C*.

Our explanation is that as the diffusivity of the molecular complexes increases, they become more likely to diffuse away from the zone of binding and eventually dissociate, resulting in less stable patterns. If the molecular complexes are as mobile as free molecules, the synapse will never form as the complexes will drift out of the zone of binding at the same rate as free molecules will drift into it. This also is the reason for the lower number of BCR/Ag complexes formed with increasing complex diffusivity.

II. Significant Membrane Deformation

*Synapses cannot form by simple diffusion and binding (No shift in LFA-1 affinity)*

Interestingly, no synapse is observed to form by the diffusion-and-binding mechanism mentioned previously when we use equilibrium BCR/Ag complex length equal to that of LFA-1/ICAM-1, ($z_{eq(BA)}$=42 nm) as is the case in *in vitro* experiments (5-7), regardless of the value of BCR affinity, BCR/Ag bond stiffness, antigen concentration or complex diffusivity. We propose that when BCR/Ag complexes and LFA-1/ICAM-1 complexes have the same length, the membrane in the region where binding is possible (Fig. 2) will be pulled down to a uniform value of *z*=42 nm, thereby resulting in a flat contact region seen in Figure 8. The effect of membrane curvature, which previously was crucial to synapse formation by allowing the BCR/Ag complexes to collect into a cluster at a faster rate than the LFA-1/ICAM-1 complexes when BCR affinity was less than LFA-1 affinity, is now entirely negated. As the membrane in the contact region assumes this flat shape rather rapidly (*t*~2000 time steps, Fig. 8), the initial ring of BCR/Ag complexes never compresses into a cluster, resulting in the purely scrambled pattern in Figure 9*A* instead of a synapse. The flattening of the membrane in the contact region also negates the previously helpful role that increased BCR/Ag bond stiffness played in synapse formation at high affinity, as the flat shape of the contact region makes differences in bond stiffness irrelevant. BCR/Ag complexes can thus form at the outer edge of the contact region, as the membrane separation distance there is the same as at the center. As the contact region flattens out rather rapidly, the ring-like pattern will persist, thereby generating the pattern seen in Figure 9*B*.

Synapses are observed to form across the entire range of BCR affinity values ($K_A$= $10^6$-$10^{10}$ M$^{-1}$), however, when we set the length of the BCR/Ag complexes to its minimum theoretically possible value, $z_{eq(BA)}$=22 nm. As before, at high affinity it is necessary for the stiffness of the BCR/Ag bond to be considerably higher than that of the LFA-1/ICAM-1 bond (~10-fold difference). With shorter BCR/Ag bond length it is possible for synapses to form because the membrane separation at center of the contact region will become uniformly equal to the shorter BCR/Ag bond length, while the separation distance at the outside of the contact region is equal to the longer LFA-1/ICAM-1 bond length. The longer LFA-1/ICAM-1 complexes are thus excluded from the synapse region, generating a synapse pattern. It also is necessary, however, for the BCR/Ag bond to be considerably stiffer than the LFA-1/ICAM-1 bond, otherwise the membrane separation distance will tend towards a uniform intermediate value, which is not conducive to synapse formation. The minimum stiffness value needed to generate a synapse in this scenario is slightly higher than for the case of no membrane deformation (second column of Table 5).

*A shift in LFA-1 affinity can drive synapse formation at low BCR affinity*

Our model shows it is possible for synapses to form with $z_{eq(BA)}$=42 nm and significant membrane deformation, however, when the affinity of LFA-1 undergoes a shift from low to high as a result of BCR/Ag binding, as outlined previously. At the lowest affinity simulated, $K_A$=$10^6$ M$^{-1}$, synapses were observed to form when the stiffness of the BCR/Ag bond was one-and-a-half that of the LFA-1/ICAM-1 bond, shown in Figure 9*C*. As affinity increased, higher values of BCR/Ag bond stiffness were necessary to produce a synapse, such that a 4-fold increase in stiffness was necessary at $K_A$=$10^7$ M$^{-1}$ and a 10-fold increase at $K_A$=$10^8$ M$^{-1}$. At the two highest affinity values simulated, $K_A$=$10^{9-10}$ M$^{-1}$, no synapses were observed for any BCR/Ag bond stiffness value, instead generating the pattern seen in Figure 9*D*.

We observe that when LFA-1 affinity is initially low, almost all complexes formed are BCR/Ag complexes. Because fewer complexes are formed, the central portion of the membrane flattens more slowly than when BCR/Ag and LFA-1/ICAM-1 complexes form simultaneously. This gives the initial ring of BCR/Ag complexes enough time to compress into a cluster before the center of the contact region flattens out. By the time the affinity of LFA-1 is shifted, allowing rapid LFA-1/ICAM-1 binding, the center of the contact zone is already occupied by the BCR/Ag complexes, so that the LFA-1/ICAM-1 complexes cannot achieve numerical superiority in the center but do so at the periphery, generating a synapse pattern such as in Fig. 9*C*. With increasing BCR affinity, it is more difficult for the initial ring of BCR/Ag complexes to compress into a cluster, so that a greater bond stiffness in needed to produce a synapse. At the highest BCR affinity values, the rate of BCR/Ag complex formation is sufficiently high that the central portion of the membrane flattens out rapidly, with the result that the ring-like pattern formed by the BCR/Ag complexes persists no matter what the BCR/Ag bond stiffness.

      As before, synapses form over the entire range of BCR affinity values ($K_A = 10^6$-$10^{10}$ $M^{-1}$), when the length of the BCR/Ag complexes is set to its minimum theoretically possible value, $z_{eq(BA)}$=22 nm. The shift in LFA-1 affinity only enhances the length-dependent synapse formation mechanism, by giving the BCR/Ag complexes more time to pull down the center of the membrane to $z$=22 nm, thereby producing a more clear demarcation between the zone where $z$=42 nm and $z$=22 nm, and thus more defined synapses.

# DISCUSSION

Our modeling results predict the existence of multiple, potentially overlapping synapse formation mechanisms in B cells. The primary determinants the particular mechanism necessary for synapse formation are affinity and membrane deformability. The requirement that low molecular complex diffusivity is crucial to synapse formation in all cases indicates these are all diffusion-and-binding type mechanisms. However, the basic diffusion-and-binding mechanism proposed by Iber (19) appears sufficient to drive synapse formation only when (i) BCR affinity is lower than LFA-1 affinity and (ii) the membrane does not appreciably change shape over the timescale of synapse formation. In the case of high BCR affinity without membrane deformation, higher BCR/Ag bond stiffness as compared to LFA-1/ICAM-1 is sufficient to drive synapse formation. With the addition of significant membrane deformation, we observe that synapse formation at low BCR affinity either requires higher BCR/Ag bond stiffness or a signaling-driven shift in the affinity of LFA-1 (i.e. basic diffusion-and-binding is not sufficient). At high BCR affinity and membrane deformation, however, our model shows only differences in length between BCR/Ag and LFA-1/ICAM-1 complexes can generate a synapse. Since BCR/Ag and LFA-1/ICAM-1 may have comparable extracellular domain length (5-7), this suggests that yet another mechanism, such as signaling-driven active transport of receptor molecules by cytoskeletal rearrangement, may be necessary for synapse formation in this regime.

Based on our model's results, we propose the following as the most probable mechanism of synapse formation in B cells: As the B cell and APC approach each other, the LFA-1 is in a low affinity state, so that binding begins when the two cells are close enough for a BCR molecule to bind with an antigen molecule. BCR molecules from the periphery drift into the zone where binding is possible, whereupon they either bind to antigen and form a complex, or eventually drift back out to the periphery again. Because molecular complexes tend to have low diffusivity, they tend to stay where they form, at the edges of the zone of binding, producing and ring-like pattern that becomes more cluster-like over time. During this process the B cell membrane deforms to accommodate the BCR/Ag complexes at their equilibrium bond length. After a critical number of BCR/antigen complexes have formed, outside-in signaling from the activated BCR results in a shift in the confirmation of LFA-1 to a high affinity state. With the addition of significant LFA-1/ICAM-1 binding, membrane deformation is accelerated so the center of the contact zone rapidly assumes a more or less flat shape. The head start in binding of the BCR, combined with the potentially higher bond stiffness and shorter length of the BCR/Ag bond, results in BCR/Ag complexes being numerically dominant at the center of the contact zone. The potentially longer and more flexible LFA-1/ICAM-1 bonds are more numerous at the outer part of the contact zone, producing the canonical immunological synapse pattern. It should here be noted that our model failed to produce synapses at high BCR affinity ($K_A > 10^8$ $M^{-1}$) when membrane deformation is significant and when there is no difference in length between BCR/Ag and LFA-1/ICAM-1 complexes. Synapse formation in such a situation might require active transport of receptors as a result of cytoskeletal rearrangement and the results of one recent experiment (9) indeed support such an idea.

Comparison of our model's result with experimental data shows substantial agreement. With the mechanism described above, our model predicts synapses in the case of stiff membranes over the entire range of physiological BCR affinity values mentioned in the literature ($K_A=10^6$-$10^{10}$ M$^{-1}$) (6,7).  Furthermore, our model does not produce synapses below $K_A=10^6$ M$^{-1}$, which is in line with experimental results (6,7).  The minimum number of antigen molecules needed for a synapse is roughly 500 molecules, which corresponds to a concentration of ~50 molecules/µm$^2$, assuming a contact area of ~9 µm$^2$.  The size of the synapses predicted by our model is around 1 µm in diameter, which is somewhat smaller than the 2 µm diameter of real synapses.  This difference is reduced, however, the length of BCR/Ag complexes is less than that of LFA-1/ICAM-1 ($z_{eq(BA)}$=22 nm), in which case the diameter of synapses predicted by our model can be as large as 1.5 µm.  The time scale of synapse formation in our model is of the order of 10$^4$ time steps, which is mapped to physical time by matching the diffusion coefficients in our simulation to those reported in experiments, thus allowing the relation between our model's time scale and physical time to emerge naturally (see Appendix).  This process yields a one centisecond-per time step mapping, which means our model's time scale of synapse formation corresponds to the experimentally observed synapse formation time of 1-2 minutes.

      As it stands, our model possesses several attributes that make it particularly suitable for modeling B cell synapse formation.  The approach we are using is stochastic and discrete in nature, and hence is suitable for the modeling of situations of low antigen concentration, such as the onset of the immune response.  In addition, we use a Monte Carlo scheme that is computationally efficient and can thus carry out an entire set of virtual experiments in a matter of minutes. Furthermore, we present a novel framework for mapping our model's probabilistic parameters into physical quantities and vice-versa (see Appendix).  Such a framework is notably absent from similar Monte Carlo models developed to study such systems in the past and to the best of our knowledge is the first of its kind.  Finally, our model is very general in nature and can potentially be used to model a variety of similar cell-cell systems.

      Nevertheless, our model in its current form can be made more physiological in a variety of ways, especially with regards to the modeling of signaling-induced processes. Most important is the addition of signaling-driven convective motion of receptors by cytoskeletal rearrangement.  Another extension of our work could involve further exploration of the form of the LFA-1 affinity shift, using time histories other than the current step function.

      One of the main goals of this study is to make predictions about the process of B cell synapse formation that may be experimentally tested.  Our model's predictions on the effect of bond stiffness, equilibrium length, membrane deformation and time history of LFA-1 affinity are particularly amenable to experimental investigation.  It is our belief the combination of computational modeling and experimental results in an iterative process can lead to a full understanding of the process of immunological synapse formation in B cells and further account for the physiological responses observed during B cell immune function.

APPENDIX

Because some of the parameters of our model are probabilistic in nature and therefore dimensionless, it is necessary to map them onto physical quantities to be able to physically interpret the results. Two such mappings are necessary: One which maps the probabilistic affinity $P_A^{max}$ to the association constant $K_A$ and one which maps the length of our model's time step, $p_{diff}$, $p_{on}^{max}$ and $p_{off}^{min}$ to physical time, the diffusion coefficient $D$, $k_{on}$, and $k_{off}$, respectively.

We begin this section with the mapping between $P_A^{max}$ and the association constant $K_A$. To map values of $P_A^{max}$ onto corresponding values of $K_A$, we make use of the fact that at kinetic equilibrium, the two-dimensional association constant, $K_{A(2D)}$, can be obtained from the following relation (12,26,29):

$$K_{A(2D)} = \frac{C_{complex}}{C_{free(1)} \cdot C_{free(2)}} = \frac{N_{complex}}{N_{free(1)} * N_{free(2)}} \cdot Area \quad (A.1)$$

Here $C$ refers to the concentration (molecules/area), $N_{complex}$ is the number of complexes formed at equilibrium, while $N_{free(1)}$ and $N_{free(2)}$ refer to the number of free molecules present at equilibrium. To map $P_A^{max}$ to $K_{A(2D)}$, we run our simulation for a particular value of $P_A^{max}$ to obtain $N_{complex}$, $N_{free(1)}$, and $N_{free(2)}$, and calculate $K_{A(2D)}$ from Eq. A.1.

The results are shown in Figure A.1, where we see a linear relationship between $K_{A(2D)}$ and $P_A^{max}$ of the form:

$$K_{A(2D)} = (2*10^{-3} \, \mu m^2/molec.) * P_A^{max} \quad (A.2)$$

Because the affinity of BCR in the experimental literature is usually given in units of 3-D $K_A$, it also is necessary to convert values of $K_{A(2D)}$ to $K_{A(3D)}$. This is done by first multiplying by the effective confinement length in the manner of Bell (29), for which we use the thickness of cell membrane (~10 nm). Since $K_{A(3D)}$ is usually given in units of $M^{-1}$, the second step in the conversion consists of multiplying by the conversion factor $1L=(0.1m)^3=10^{15}\mu m^3$ and multiplying by Avogadro's number (1 mol=6*10^{23} molecules). This results in the following relation between $P_A^{max}$ and $K_{A(3D)}$:

$$K_{A(3D)} = \left(\frac{2*10^{-3} \, \mu m^2}{molec.}\right) * P_A^{max} * 0.01 \, \mu m * \frac{1\,L}{10^{15}\,\mu m^3} * \frac{6*10^{23}\,molec.}{1\,mole} = (10^4 M^{-1}) * P_A^{max} \quad (A.3)$$

Thus, for example, the reported value of LFA-1 affinity of 3.3 $\mu m^2$/molecule (28) approximately maps to $P_{A(LI)}^{max}$=1000 (using Eq. A.2), which in turn corresponds to $K_{A(3D)}=10^7$ $M^{-1}$ (using Eq. A.3).

Next, we establish the mapping of our model's time scale to physical time. There are two ways of doing this: One is to match the number of time steps it takes to obtain a synapse in our model to the time scale of synapse formation in experiments, and from there map $p_{diff}$, $p_{on}^{max}$ and $p_{off}^{min}$ to their physical counterparts. Another is to match the $p_{diff}$ for which we obtain a synapse in our model to the diffusion $D$ reported in physical experiments, and allow the time scale of our model to emerge naturally from this. Because it appears more sound, we use the latter approach.

As with affinity, we map the probability of diffusion $p_{diff}$ to the diffusion coefficient $D$ by means of direct simulation. In these simulations, we note the location and time of each molecule as it is created. For complexes, this is simply the time and location at which they form, while for free molecules this is either their initial location on the lattice and $t=0$, or if they have been created as a result of a complex dissociating, the location and time at which the complex dissociated. At each time step, the square of the distance the molecule has traveled from its location of creation is divided by the number of time steps that molecule has been in existence. This is then averaged over all the molecules of that particular type to obtain the simulation diffusion coefficient, $D_{sim}$, at that particular time step. Thus we have:

$$D_{sim} = \frac{\sum_{i=1}^{N} \frac{(x_{curr} - x_o)^2 + (y_{curr} - y_o)^2}{(t_{curr} - t_o)}}{N} \quad (A.4)$$

We then run the simulation for a particular value of $p_{diff}$ to obtain a time plot of the value of $D_{sim}$ in the manner of Figure A.2. From the figure, we see that a probability of diffusion $p_{diff}=1$ corresponds to a value of $D_{sim}$ in the range of 0.1-1 (nodal spacings)$^2$/time step. We then multiply by the appropriate conversion factor to convert the length in nodes to physical length:

$$1.0 * \frac{(\text{nodal spacings})^2}{\text{time step}} * \frac{(0.01 \mu m)^2}{(1 \text{ nodal spacing})^2} = 10^{-4} \frac{\mu m^2}{\text{time step}} \quad (A.5)$$

We now match this value to that of diffusion coefficient in synapse experiments found in the literature to obtain the physical length of time of one of our model's time steps. The literature value of the diffusion coefficient in synapse experiments of ~0.01 $\mu m^2$/sec indicates that a single time step in our model corresponds to 0.01 seconds, i.e. a one centisecond-per time step mapping. The observed time of synapse formation of $t=10^4$ time steps in our simulations thus corresponds to 100 seconds, which agrees rather well with the experimental time of synapse formation of 1-2 minutes. With this time scale mapping, the diffusion coefficient mapping now becomes:

$$D_{phys} = \frac{(0.01 \mu m)^2}{(1 \text{ nodal spacing})^2} * \frac{10^2 \text{ time steps}}{1 \text{ second}} * D_{sim} = 0.01 \frac{\mu m^2}{\text{sec}} * D_{sim} \quad (A.6)$$

Once we have obtained the time scale mapping, it is straightforward to map $p_{off}^{min}$ to $k_{off}$ through the relation:

$$k_{off} = \frac{10^2 \text{ time steps}}{\text{sec}} \cdot p_{off}^{min} \quad (A.7)$$

Thus, the reported $k_{off}$ for LFA-1 in the literature of 0.1 s$^{-1}$ (28) corresponds to $p_{off(LI)}^{min}=10^{-3}$. Multiplying Eqs (A.3) and (A.7), we obtain the mapping between $p_{on}^{max}$ and $k_{on}$:

$$k_{on} = (10^6 \text{ M}^{-1}\text{s}^{-1}) * p_{on}^{max} \quad (A.8)$$

From this, we estimate the measured value of $k_{on}=2*10^6$ M$^{-1}$s$^{-1}$ for the HEL line of antigens in (6,7) approximately corresponds to a $p_{on}^{max}=1.0$. To simulate values of $k_{on}$ greater than $2*10^6$ M$^{-1}$s$^{-1}$, the mapping between $p_{diff}$ and $D$ would have to be changed by matching $p_{diff}=1$ to a higher $D$ value.

ACKNOWLEDGEMENTS

The authors would like to thank Drs. Arup Chakraborty, Rajiv Singh and Volkmar Heinrich for many insightful discussions over the course of this study. We also would also like to acknowledge Dr. Roger Zauel for his valuable help on random number generation. N.B. acknowledges support from NIH grant R01 AI051354.

REFERENCES

1. Wulfing, C., M.D. Sjaastad, and M.M. Davis. 1998. Visualizing the dynamics of T cell activation: Intracellular adhesion molecule 1 migrates rapidly to the T cell/B cell interface and acts to sustain calcium levels. Proc. Natl. Acad. Sci. USA. 95:6302-6307.

2. Monks, C.R., B.A. Freiberg, H. Kupfer, N. Sciaky, and A. Kupfer. 1998. Three-dimensional segregation of supramolecular activation clusters in T cells. Nature 395:82-86.

3. Grakoui, A., S.K. Bromley, C. Sumen, M.M. Davis, A.S. Shaw, P.M. Allen, and M.L. Dustin. 1999. The immunological synapse: A molecular machine controlling T cell activation. Science 285:221-227.

4. Krummel, M.F., M.D. Sjaastad, C. Wulfing, and M.M. Davis. 2000. Differential clustering of CD4 and CD3$\zeta$ During T Cell Recognition. Science 289:1349-1352.

5. Batista, F.D., D. Iber, and M.S. Neuberger. 2001. B cells acquire antigen from target cells after synapse formation. Nature 411:489-494.

6. Carrasco, Y.R., S.J. Fleire, T. Cameron, M.L. Dustin, and F.D. Batista. 2004. LFA-1/ICAM-1 interaction lowers the threshold of B cell activation by facilitating B cell adhesion and synapse formation. Immunity 20:589-599.

7. Carrasco, Y., and F.D. Batista. 2006. B-cell activation by membrane-bound antigens is facilitated by the interaction of VLA-4 with VCAM-1. EMBO J. 25:889-899.

8. Fleire, S.J., J.P. Goldman, Y.R. Carrasco, M. Weber, D. Bray, and F.D. Batista. 2006. B cell ligand discrimination through a spreading and contracting response. *Science* 312:738-741.

9. Carrasco, Y.R., and Batista, F.D. 2006. B cell recognition of membrane-bound antigen: An exquisite way of sensing ligands. Curr. Op. Immun. 18:286-291.

10. Bray, D., M.D. Levin, and C.J. Morton-Firth. 1998. Receptor clustering as a cellular mechanism to control sensitivity. Nature 393:85-88.

11. Qi, S.Y., J.T. Groves, and A.K. Chakraborty. 2001. Synaptic pattern formation during cellular recognition. Proc. Natl. Acad. Sci. USA. 98:6548-6553.

12. Burroughs, N.J., and C. Wülfing. 2002. Differential segregation in a cell-cell contact interface: The dynamics of the immunological synapse. Biophys. J. 83:1784-1796.

13. Lee, K.H., A.R. Dinner, C. Tu, G. Campi, S. Raychaudhri, R. Varma, T.N. Sims, W.R. Burack, H. Wu, J. Wang, O. Kanagawa, M. Markiewicz, P.M. Allen, M.L. Dustin, A.K. Chakraborty, and A.S. Shaw. 2003. The immunological synapse balances T cell receptor signaling and degradation. Science 302:1218-1222.

14. Lee, S.J.E., Y. Hori, J.T. Groves, M.L. Dustin, and A.K. Chakraborty. 2002. Correlation of a dynamic model for immunological synapse formation with effector functions: two pathways to synapse formation. TRENDS Immunol. 23:492-502.

15. Lee, S.J.E., Y. Hori, and A.K. Chakraborty. 2003. Low T cell receptor expression and thermal fluctuations contribute to formation of dynamic multifocal synapses in thymocytes. Proc. Nat. Acad. Sci. *USA* 100:4383-4388.

16. Raychaudhuri, S., A.K. Chakraborty, and M. Kardar. 2003. Effective membrane model of the immunological synapse. Phys. Rev. Lett. 91:(208101-1)-(208101-4).

17. Coombs, D., M. Dembo, C. Wofsy, and B. Goldstein. 2004. Equilibrium thermodynamics of cell-cell adhesion mediated by multiple ligand-receptor pairs. Biophys. J. 86:1408-1423.

18. Weikl, T.R., and R. Lipowsky. 2004. Pattern formation during T-cell adhesion. Biophys. J. 87:3665-3678.

19. Iber, D. 2005. Formation of the B-cell synapse: Retention or recruitment? Cell. And Mol. Life Sci. 62:206-213.

20. Van Kampen, N.G. 2001. Stochastic Processes in Physics and Chemistry. Elsevier, Amsterdam.

21. Hammer, D., and Apte, S. 1991. Simulation of cell rolling and adhesion of surfaces in shear flow: General results and analysis of selectin-mediated neutrophil adhesion. Biophys. J. 63:35-57.

22. Mahama, P.A., and Linderman, J.J. 1995. Monte Carlo simulations of membrane signal transduction events: Effect of receptor blockers on G-protein activation. Ann. Biomed. Eng. 23:299-307

23. Chakraborty, A.K., M.L. Dustin, and A.S. Shaw. 2003. *In silico* models for cellular and molecular immunology: Successes, promises, and challenges. Nat. Immunol. 4:933-936.

24. Goldstein, B., J.R. Faeder, and W.S. Hlavacek. 2004. Mathematical and computational models of immune-receptor signaling. Nat. Rev. Immunol. 4:445-456.

25. Dembo, M., T.C. Torney, K. Saxman, and D. Hammer. 1988. The reaction-limited kinetics of membrane-to-surface adhesion and detachment. Proc. R. Soc. Lond. B 234:55-83.

26. Lauffenburger, D.A., and J.J. Linderman. 1993. Models for binding, trafficking and

signaling. Oxford University Press, Oxford.

27. Tominaga, Y., Y. Kita, A. Satoh, S. Asai, K. Kato, K. Ishikawa, T. Horiuchi, and T. Takashi. 1998. Expression of a soluble form of LFA-1 and demonstration of its binding activity with ICAM-1. J. Immunol. Meth. 212:61-68.

28. Favier, B., N.J. Burroughs, L. Weddeburn, S. Valitutti. 2001. T cell antigen receptor dynamics on the surface of living cells. Int. Immunol. 13:1525-1532.

29. Bell, G.I. 1983. Cell-cell adhesion in the immune system. Immunol. Today 4:237-240.

30. http://www.accessexcellence.org/RC/VL/GG/antiBD_mol.html

31. Alberts, B., Johnson, A., Lewis, J., Raff, M., Roberts, K., and Walter, P. 2002 Molecular biology of the cell, 4$^{th}$ ed. Garland Science, London. p. 1375.

32. Lanzavecchia, A. 1985. Antigen-specific interaction between T and B cells. Nature 314:537-539.

33. Batista, F.D., and M.S. Neuberger. 1998. Affinity dependence of the B-cell response to antigen: A threshold, a ceiling, and the importance of off-rate. Immunity 8:751-759.

34. Kouskoff, V., S. Famiglietti, G. Lacaud, P. Lang, J.E. Rider, B.K. Kay, J.C. Cambier, and D. Nemazee. 1998. Antigens Varying in Affinity for the B Cell Receptor Induce Differential B Lymphocyte Responses. J. Exp. Med. 188:1453-1464.

35. Shih, T.A., E. Meffre, M. Roederer, and M.C. Nussenzweig. 2002a. Role of antigen receptor affinity in T cell-independent antibody responses in vivo. Nat. Immunol. 3:399-406.

36. Shih, T.A., E. Meffre, M. Roederer, and M.C. Nussenzweig. 2002b. Role of BCR affinity in T cell dependent responses in vivo. Nat. Immunol. 3:570-575.

**Table 1.** Spatial dimensions of the model.

| Parameter | Value |
| --- | --- |
| Size of contact region | 3μmX3μm |
| Number of nodes | 300X300 |
| Nodal spacing | 10 nm |
| Cell radius (B Cell, APC) | 6 μm |
| $z_{max}$ from eq. (1) (cell-bilayer case) | 350 nm + $z_o$ |

**Table 2.** Parameters of the model.

| Parameter | Description |
| --- | --- |
| $p_{on(BA)}^{max}$ | Maximum BCR/Ag complex formation probability |
| $p_{off(BA)}^{min}$ | Minimum BCR/Ag complex dissociation probability |
| $p_{on(LI)}^{max}$ | Maximum BCR/Ag complex formation probability |
| $p_{off(LI)}^{min}$ | Minimum LFA1/ICAM1 complex dissociation probability |
| $B_0$ | Initial number of free BCR molecules |
| $A_0$ | Initial number of free antigen molecules |
| $L_0$ | Initial number of free LFA-1 molecules |
| $I_0$ | Initial number of free ICAM-1 molecules |
| $\kappa_{BA}$ | Stiffness of BCR/Ag bond |
| $\kappa_{LI}$ | Stiffness of LFA-1/ICAM-1 bond |
| $z_{eq(BA)}$ | Equilibrium extracellular length of BCR/Ag complex |
| $z_{eq(LI)}$ | Equilibrium extracellular length of LFA-1/ICAM-1 complex |
| $p_{diff(B)}$ | Probability of diffusion of a free BCR molecule |
| $p_{diff(A)}$ | Probability of diffusion of a free antigen molecule |
| $p_{diff(L)}$ | Probability of diffusion of a free LFA-1 molecule |
| $p_{diff(I)}$ | Probability of diffusion of a free ICAM-1 molecule |
| $p_{diff(BA)}$ | Probability of diffusion of a BCR/Ag complex |
| $p_{diff(LI)}$ | Probability of diffusion of a LFA1/ICAM1 complex |
| $M$ | Time scale of membrane deformation |
| $\gamma$ | Membrane tension |
| $\beta$ | Membrane bending rigidity |

**Table 3.** Experimentally measured parameter values and their probabilistic counterparts.

| Experimental Parameter | Measured Value | Simulation Parameter | Mapped Value |
|---|---|---|---|
| $K_A$ BCR/Ag | $10^6$-$10^{10}$ M$^{-1}$ (6,7) | $P_{A(BA)}^{max}$ | $10^2$-$10^6$ |
| $k_{on}$ BCR/Ag | $10^4$-$10^6$ M$^{-1}$s$^{-1}$ (6,7) | $p_{on(BA)}^{max}$ | 0.01-1 |
| $k_{off}$ BCR/Ag | 1-$10^{-4}$ s$^{-1}$ (6,7) | $p_{off(BA)}^{min}$ | $10^{-2}$-$10^{-6}$ |
| $K_A$ LFA-1/ICAM-1 | 3.3 µm$^2$/molec. (28) | $P_{A(LI)}^{max}$ | $10^3$ |
| $k_{on}$ LFA-1/ICAM-1 | 0.33 µm$^2$·s$^{-1}$/molec. (28) | $p_{on(LI)}^{max}$ | 1.0 |
| $k_{off}$ LFA-1/ICAM-1 | 0.1 s$^{-1}$ (28) | $p_{off(LI)}^{min}$ | $10^{-3}$ |
| Antigen conc. | 10-1000 molec./µm$^2$ (6) | $A_0$ | 100-10,000 molec. |
| ICAM-1 conc. | 170 molec./µm$^2$ (6) | $I_0$ | ~2000 molec. |
| $\kappa_{LI}$ | 40 µN/m (12) | $\kappa_{LI}$ | same |
| $z_{eq(LI)}$ | 42 nm (12) | $z_{eq(LI)}$ | same |
| $D$ free molec. | ~0.01 µm$^2$/sec (27) | $p_{diff(F)}$ | 1.0 |
| $\gamma$ | 24 µN/m (12) | $\gamma$ | same |
| $\beta$ | 5*$10^{-20}$ J (12) | $\beta$ | same |

**Table 4.** Unknown or variable parameters.

| Parameter | Type |
|---|---|
| $p_{on(BA)}^{max}$ | Known, variable |
| $p_{off(BA)}^{min}$ | Known, variable |
| $A_0$ | Known, variable |
| $\kappa_{BA}$ | Unknown, may vary |
| $z_{eq(BA)}$ | May vary between ~22-42 nm |
| $M$ | Unknown |

**Table 5.** Minimum $\kappa_{BA}$ (µN/m) value needed for synapse formation.

| BCR Affinity $K_{A(BA)}$ (M$^{-1}$) | $z_{(eq)BA}$=42nm | $z_{(eq)BA}$=22nm |
|---|---|---|
| $10^{10}$ | 400 | 320 |
| $10^9$ | 320 | 280 |
| $10^8$ | 240 | 200 |
| $10^7$ | 160 | 120 |
| $10^6$ | 40 | 40 |

Figure 1. Model of the B cell-APC contact region. The cells are assumed to have a spherical shape, with the total vertical separation distance between the two surfaces at any point $z=z_1+z_2$, with $z_1$ and $z_2$ given by Eq. 1. At the center of the contact zone ($x_o$, $y_o$), the vertical separation distance is at its minimum $z=z_o$, while at the corners it is $z=zmax$. The 3µm X 3µm simulated area is large enough to include the entire zone where binding is possible.

Figure 2. Sample graphical representation of (A) $p_{on}$, (B) $p_{off}$, and (C) $P_A$ according to Eqs. 2-4. Receptor-ligand binding can only occur where $p_{on}^{max}>0$ and $p_{off}^{min}<1$. In this set of images, $\kappa'=40$ µN/m, $\kappa=2\kappa'$, $z_{eq}=42$ nm, $k_B=1.38*10^{-23}$ J/K, $T=300$ K, $p_{on}^{max}=1.0$ and $p_{off}^{min}=0.001$.

Figure 3. Monte Carlo algorithm flow chart.

Figure 4. Effect of varying BCR affinity ($P_{A(BA)}^{max}$) on synapse formation. BCR/Ag complexes are shown in green, while LFA-1/ICAM-1 complexes are shown in red. The complexes are plotted in random order so as to simulate experimental intensity plots as closely as possible. In this set of figures the affinity of BCR was varied from $P_{A(BA)}^{max}=10$ to $P_{A(BA)}^{max}=10^4$ ($K_A \approx 10^5$-$10^8$ M$^{-1}$). In 4(A) the affinity is too low for synapse formation but in Fig. 4(B) we see that the difference in affinity between BCR and LFA-1 is sufficient to produce a synapse. This is no longer the case in 4(C), where the affinities are equal, while an inverted pattern forms in 4(D), where $P_{A(BA)}^{max}>P_{A(LI)}^{max}$. These images were taken after $T=10^4$ time steps (100 s) with $P_{A(LI)}^{max}=1000$ ($K_A \approx 10^7$ M$^{-1}$), $A_0=I_0=2000$ molecules, $\kappa_{BA}=\kappa_{LI}=40$ µN/m, $z_{eq(BA)}=z_{eq(LI)}=42$ nm, $p_{diff(F)}=1$ and $p_{diff(C)}=0.01$.

Figure 5. Minimum BCR/Ag bond stiffness (µN/m) needed to form a synapse with increasing BCR affinity when there is no membrane deformation. The stiffness of the LFA-1/ICAM-1 bond is fixed at 40 µN/m.

Figure 6. Effect of varying initial antigen concentration ($A_0$) on synapse formation. In this set of images $A_0$ was varied from 2000 to 200 molecules. No synapse was seen to form below $A_0=500$ molecules, while there is little change in the synapse pattern above this value. This set of images was obtained after $T=10^4$ time steps (100 s) with $P_{A(BA)}^{max}=100$ ($K_A \approx 10^6$ M$^{-1}$), $P_{A(LI)}^{max}=1000$ ($K_A \approx 10^7$ M$^{-1}$), $I_0=2000$ molecules, $\kappa_{BA}=\kappa_{LI}=40$ µN/m, $z_{eq(BA)}=z_{eq(LI)}=42$ nm, $p_{diff(F)}=1$ and $p_{diff(C)}=0.01$.

Figure 7. Effect of increasing complex diffusivity on synapse formation. The quality of the synapse seen to form when (A) $p_{diff(F)}=1$ and $p_{diff(C)}=0.01$ deteriorated significantly when the difference in diffusivity was reduced to (B) $p_{diff(F)}=1$ and $p_{diff(C)}=0.1$, and still further when (C) $p_{diff(F)}=p_{diff(C)}=1$. This dependence on complex diffusivity was observed across all synapse formation mechanisms, irrespective of all other parameter values. These images were taken after $T=10^4$ time steps (100 s) with $P_{A(BA)}^{max}=10^6$ ($K_A \approx 10^{10}$ M$^{-1}$), $P_{A(LI)}^{max}=1000$ ($K_A \approx 10^7$ M$^{-1}$), $A_0=1000$, $I_0=2000$ molecules, $\kappa_{BA}=400$ µN/m, $\kappa_{LI}=40$ µN/m, $z_{eq(BA)}=z_{eq(LI)}=42$ nm.

Figure 8. Cross-sectional view of the time evolution of membrane deformation as a result of receptor-ligand binding. Already by $t=20$ s, we can see that the center of the contact zone has more or less flattened to a separation distance $z\sim40$ nm. This image sequence was obtained with $P_{A(BA)}^{max}=100$ ($K_A\approx10^6$ M$^{-1}$), $P_{A(LI)}^{max}=1000$ ($K_A\approx10^7$ M$^{-1}$), $A_0=l_0=2000$ molecules, $\kappa_{BA}=\kappa_{LI}=40$ µN/m, $z_{eq(BA)}=z_{eq(LI)}=42$ nm, $p_{diff(F)}=1$, $p_{diff(C)}=0.01$, $M\sim10^{-12}$ m$^4$/Js, $\gamma=24$ µN/m, $\beta=5*10^{-20}$ J.

Figure 9. Effect of membrane deformation and LFA-1 affinity shift on synapse formation when $z_{eq(BA)}=z_{eq(LI)}=42$ nm. In (A) and (B) no synapse pattern is observed to form when LFA-1 is in a high affinity state from the start and membrane deformation is significant (Fig. 10). In (C), by contrast, we see a synapse form at low BCR affinity ($K_A\approx10^6$ M$^{-1}$) when LFA-1 is initially in a low affinity state and switches to the high affinity state after ~100 BCR/Ag complexes have formed ($t\sim50$ sec). In (D), we see that at high affinity this effect is lost, so the same inverted synapse pattern form as in (B). These images were taken after $T=10^4$ time steps (100 s) with $P_{A(LI)}^{max}=1000$ ($K_A\approx10^7$ M$^{-1}$), $A_0=l_0=2000$ molecules, $\kappa_{LI}=40$ µN/m, $z_{eq(BA)}=z_{eq(LI)}=42$ nm, $p_{diff(F)}=1$ and $p_{diff(C)}=0.01$, $M=10^4\ 2.4*10^{-12}$ m$^4$/Js, $\gamma=24$ µN/m, and $\beta=5*10^{-20}$ J.

Figure A.1. Mapping between simulated affinity $P_A^{max}$ and $K_{A(2D)}$. The relationship is perfectly linear, with order-of-magnitude increases in $P_A^{max}$ corresponding to order of magnitude increases in $K_{A(2D)}$.

Figure A2. Plot of $D_{sim}$ resulting from setting $p_{diff}=1.0$.

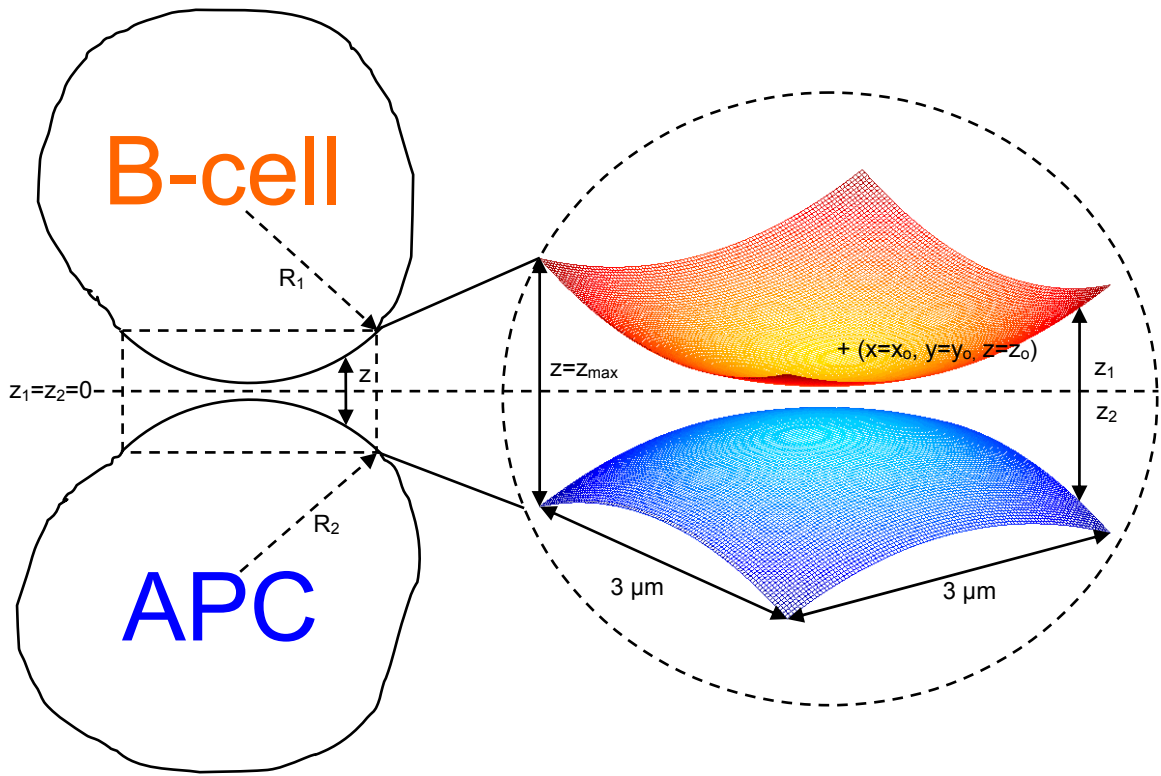

Figure 1.

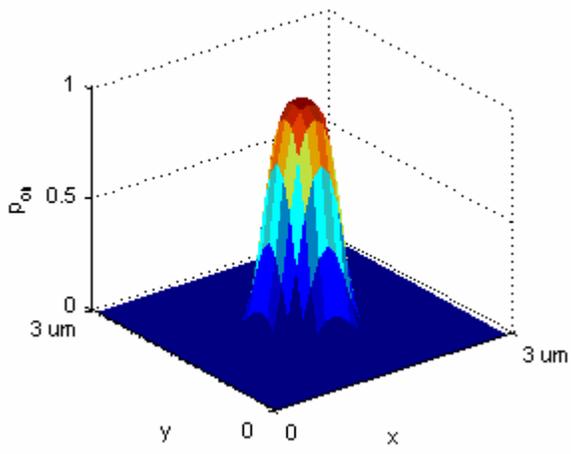

A

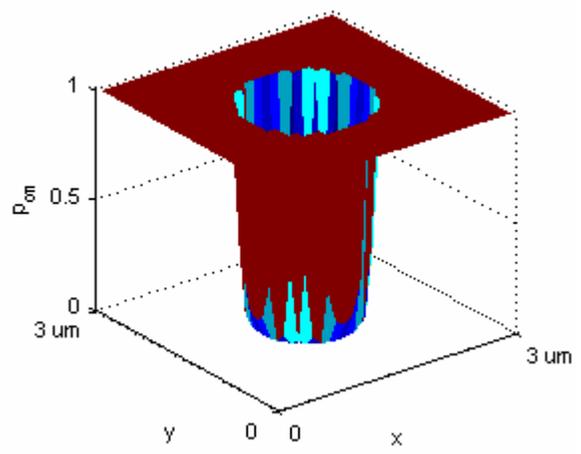

B

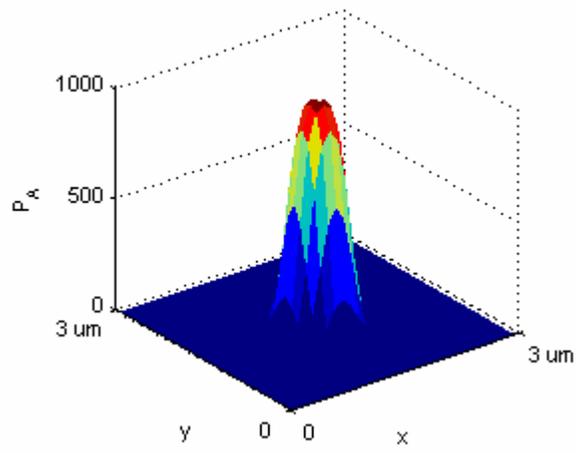

C

Figure 2.

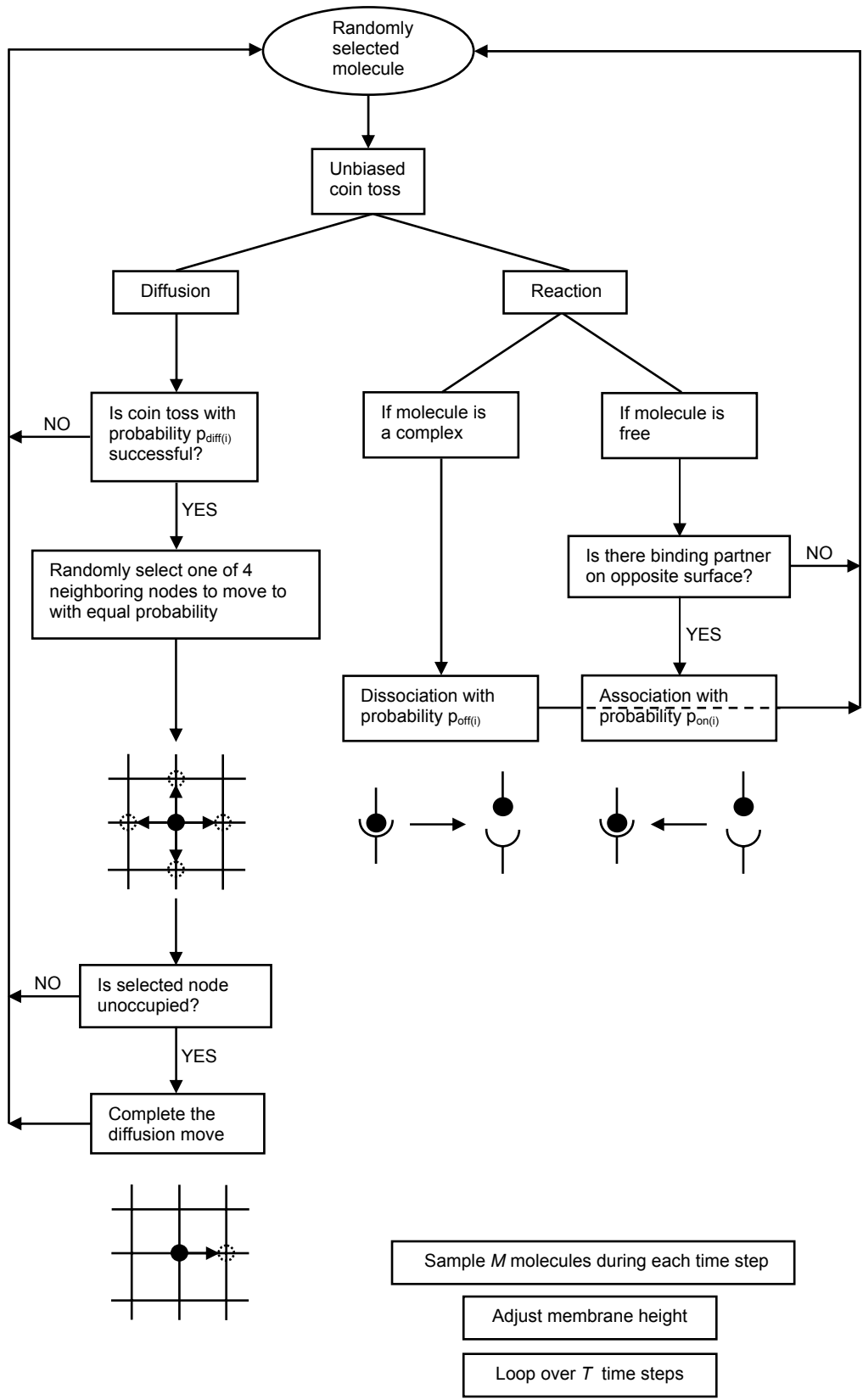

Figure 3.

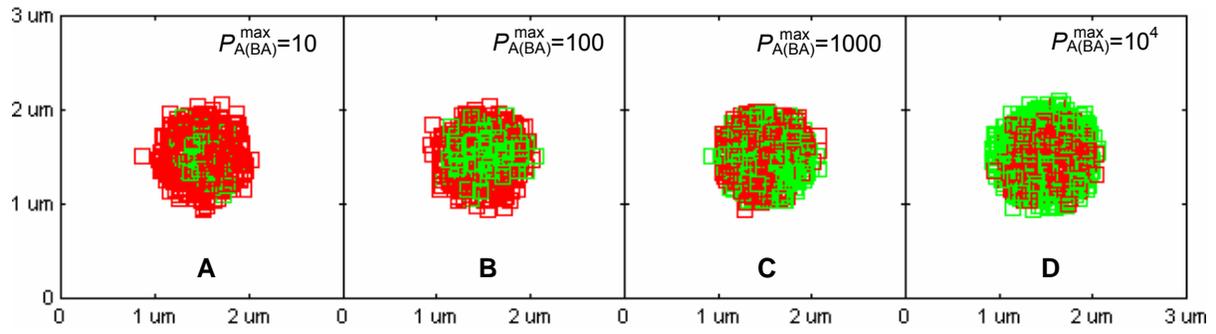

Figure 4.

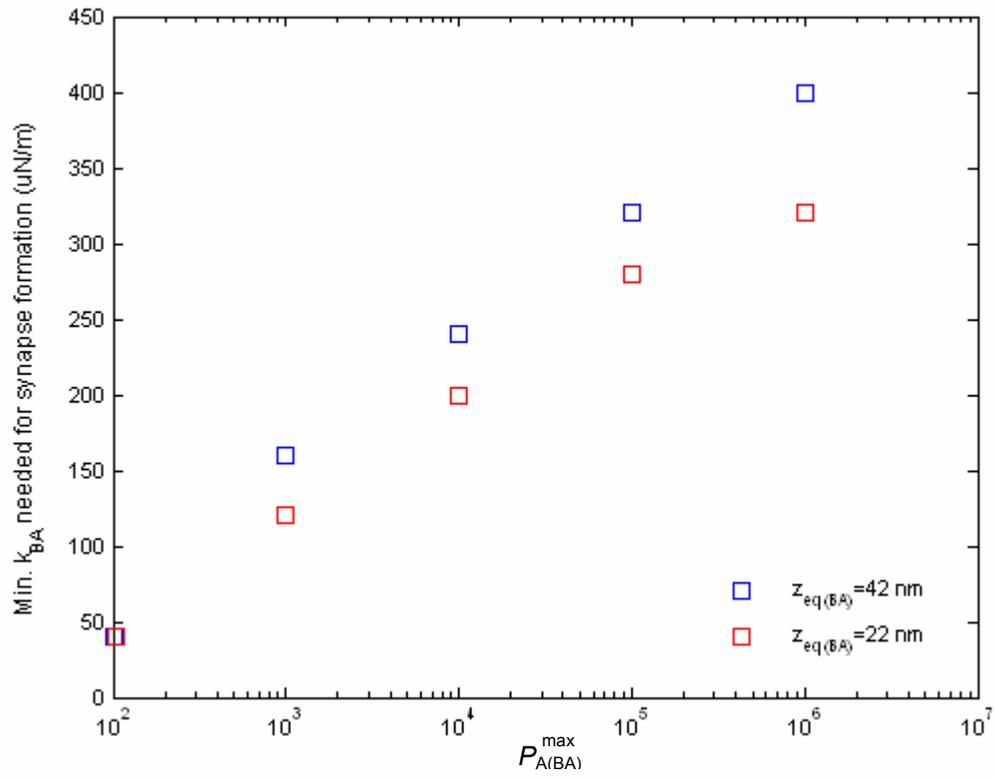

Figure 5.

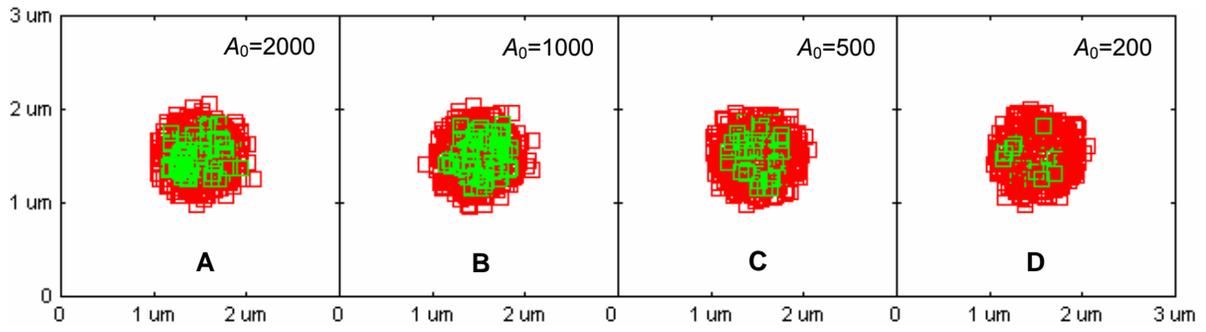

Figure 6.

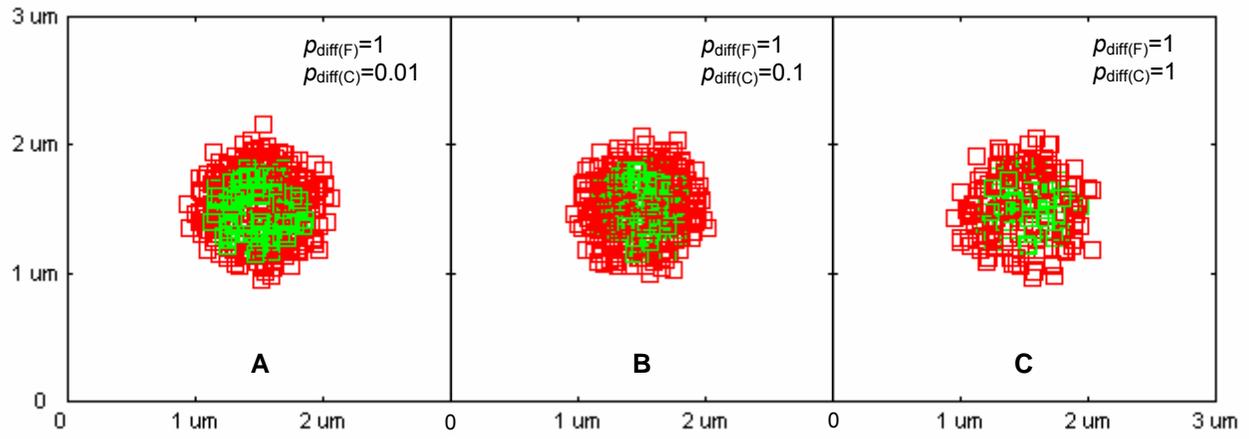

Figure 7.

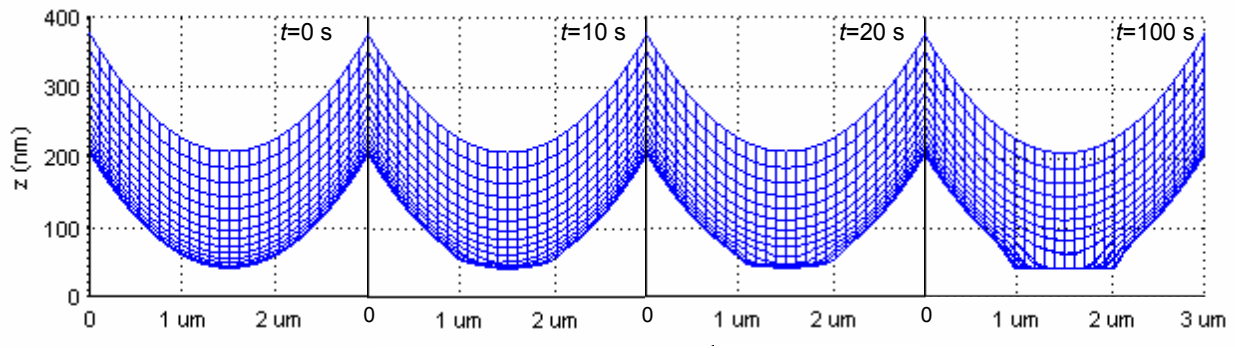

Figure 8

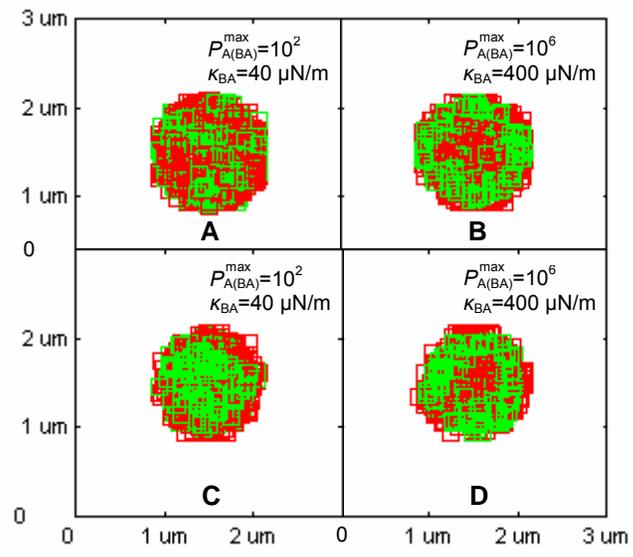

Figure 9.

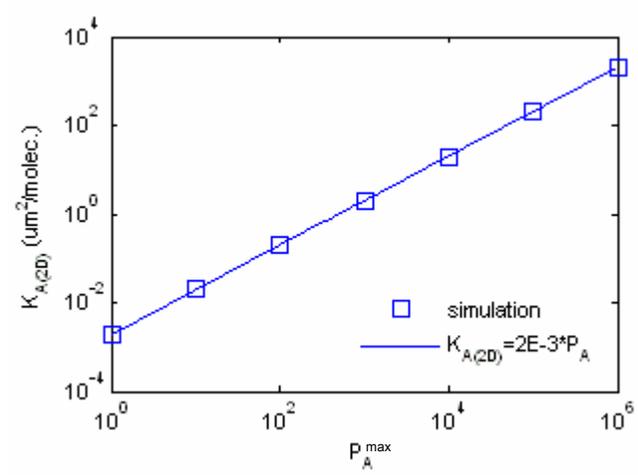

Figure A1.

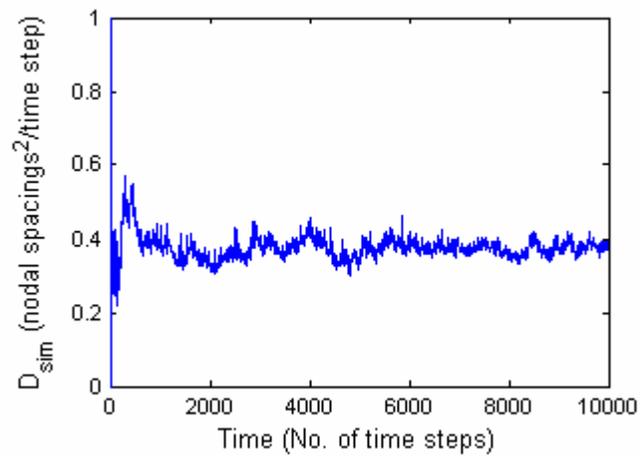

Figure A2.

## Supplemental Materials

Time scale and thermodynamic analysis of synapse formation

In this section we discuss the evolution of the synapse patterns formed in Figure 4, which are due to the difference in affinity between BCR and LFA-1. In Figure S.1 and S.2, we show the time evolution of the pattern in Figure 4$B$ and 4$D$, respectively. In both cases, the affinity of LFA-1 is kept constant at $P_{A(LI)}^{max}=1000$ ($K_A \approx 10^7$ M$^{-1}$), with $p_{on(LI)}^{max}=1.0$ and $p_{off(LI)}^{min}=0.001$. In the case of Figure S.1, the affinity of BCR is $P_{A(BA)}^{max}=100$ ($K_A \approx 10^6$ M$^{-1}$) with $p_{on(BA)}^{max}=1$ and $p_{off(BA)}^{min}=0.01$, so that we have $p_{off(BA)}^{min}>p_{off(LI)}^{min}$, which implies that the BCR/Ag complexes break up faster than the LFA-1/ICAM-1 complexes. The ring formed by the BCR/Ag complexes thus collects into a cluster faster than the ring of LFA-1/ICAM-1 complexes, producing the immunological synapse pattern seen in Figures 4$B$. In Figures S.2, BCR affinity is $P_{A(BA)}^{max}=10^4$ ($K_A \approx 10^8$ M$^{-1}$) with $p_{on(BA)}^{max}=1$ and $p_{off(BA)}^{min}=10^{-4}$, so that the situation is reversed with $p_{off(BA)}^{min}<p_{off(LI)}^{min}$, resulting in an inverted pattern. In the case when $p_{off(BA)}^{min}=p_{off(LI)}^{min}$, the pattern is purely random as seen in Figure 4$C$.

Interestingly, both patterns shown in Figures 5 and 6 are transient. At steady-state, the high affinity species, which in the Figures 5 and 6 is on the outside, will have moved to the center. Provided there are enough molecules of the higher affinity species, they eventually will completely displace the lower affinity species from the zone where binding is possible. The time to reach this steady-state, however, is estimated to be of the order of millions, if not billions of simulation time steps (~days), which is well past the biologically relevant times of 30 min-1hr (6,7). This type of slow relaxation has been observed in a variety of "glassy" systems, where very similar kinetic trapping of particles in finite regions is responsible for such long relaxation times to equilibrium. The equilibrium synapse configuration can be analyzed by a simple thermodynamic model (see below) whereas the kinetic trapping and slow relaxation in approach to equilibrium is suitably captured by our stochastic simulation.

The energy associated with receptor-ligand bond formation is related to the kinetic rate constants through the relation:

$$K_A = \frac{k_{on}}{k_{off}} = ne^{\frac{\Delta E}{k_B T}} = ne^{-E_0 + \frac{\hat{e}(z-z_{eq})^2}{2k_B T}} = ne^{-\frac{E_0}{k_B T}} e^{-\frac{\hat{e}(z-z_{eq})^2}{2k_B T}} = k_A^0 e^{-\frac{\hat{e}(z-z_{eq})^2}{2k_B T}} \quad (S.1)$$

In our stochastic model, the analog of the above equation is given by:

$$P_{A(i)}(z) = P_{A(i)}^{max} \exp\left(-\frac{\kappa_i (z-z_{eq(i)})^2}{2k_B T}\right) \quad (S.2)$$

We have established a linear mapping (see Appendix) between the association constant $K_A$ and the stochastic parameter $P_A$ associated with the probability of bond formation. This allows us to consider the equilibrium thermodynamics in terms of the stochastic parameter $P_A$.

The formation of the immunological synapse is primarily driven by the differential (spatially dependent) energy changes associated with different types of bond formation. Let us first consider for simplicity a system consisting of a single species (say BCR/Ag). We assumed in our model the closest apposition of the membranes of two spherical cells is at the center of the intercellular junction (see Fig. 1) where the separation distance is equal to the equilibrium bond length of the BCR/Ag complexes ($z_{eq(BA)}$). Clearly, the energy minimum is at the center of the intercellular junction and the equilibrium configuration (number of complexes formed as well as the spatial distribution) is governed by the Boltzmann distribution given in Eq. S.1. Interestingly, this Boltzmann distribution has a characteristic spatial dependence due to the assumed spherical shape (radius $R$) of the cells (with $z$ taken from Eq. 1):

$$P_A = P_A^{max} e^{-\frac{\kappa(z-z_{eq})^2}{2k_BT}} = P_A^{max} e^{-\frac{\kappa\left(R-\sqrt{R^2-(x^2+y^2)}\right)^2}{2k_BT}} = P_A^{max} e^{-\frac{\kappa\left(R-\sqrt{R^2-r^2}\right)^2}{2k_BT}} \quad (S.3)$$

The radius up to which bond formation will be significant is given by $P_A(r) \sim 1$. This yields the typical size of the synaptic cluster, given by the following expression:

$$r_{synapse}^2 \sim \sqrt{\frac{8k_BT(\ln P_A)R^2)}{\kappa}} \quad (S.4)$$

For a finite-sized synapse, an entropic factor arising from a larger available area as the molecules move away from the center will compete with the energy considerations arising from Eq. S.1. The radius $r$ ($<r_{synapse}$) for which the probability of occupation is highest is estimated by minimizing the free energy. The total free energy of the system is estimated as:

$$F(r) = U - TS$$

$$= -F(z_{eq}) + \frac{\kappa}{2}(z(r) - z_{eq})^2 - k_BT ln(2\pi r)$$

$$= -F(z_{eq}) + \frac{\kappa}{2}\left(R - \sqrt{R^2 - r^2}\right)^2 - k_BT ln(2\pi r) \quad (S.5\text{-}S.7)$$

Searching for the maximum by setting $\partial F/\partial r = 0$, we obtain an algebraic equation of the form:

$$r^6 + 2\alpha r^4 + (\alpha - 2\alpha R^2)r^2 - \alpha^2 R^2 = 0 \quad (S.8)$$

where $\_ = k_BT/\_$. The above equation can be solved approximately to determine $r_{max}$, the radius at which the complex molecule clustering is maximal:

$$r_{max} \sim \sqrt[4]{2}\alpha^{1/4}R^{1/2} \quad (S.9)$$

Though this type of equilibrium analysis is insightful, how the equilibrium configuration is attained (if at all) in such a reaction-diffusion system cannot be understood from such an analysis. Our simulation clearly shows that high affinity ($P_A^{max}>>1$ and $r_{synapse}\sim R$) BCR/Ag complexes are kinetically trapped far from the center, and it may take a long time to reach the final steady state central clustering. This kinetic trapping becomes even more significant when other molecular species are present.

If we include a second species in this analysis, either as free molecules or receptor-ligand complexes, then the above scenario is modified. For free molecules (with $P_A^{max} = 0$), there will be no complex formation, and hence the energy landscape will be uniform over the entire contact region. The free molecules will be eventually pushed out of the synaptic zone due to energy-reducing BCR/Ag complex formation. This rate of exclusion will crucially depend on the number of BCR/Ag bonds formed and the excluded volume effects. In our simulations no two molecules can come closer than a radius $r$ that is fixed by the nodal spacing (10 nm) of the underlying square lattice.

The most interesting case involves the addition of at least one more receptor-ligand pair besides BCR and antigen. The adhesive integrins LFA-1 and ICAM-1 are the most important molecules in this case and can be taken as a prototype for other such molecules. In terms of the affinity, we can think of two situations:
(i) $P_{A(BA)}^{max} < P_{A(LI)}^{max}$

The equilibrium configuration will consist of LFA-1/ICAM-1 complexes at the center because they have higher affinity than BCR/Ag complexes. BCR and antigen will either cluster in a ring-shaped region surrounding the LFA-1/ICAM-1 complexes, or, provided there are enough LFA-1 and ICAM-1 molecules, they will be completely excluded from the region where binding is possible. However, our simulation results clearly show that the transient BCR/Ag clustering at the center persists for a long-time before the higher affinity LFA1/ICAM-1 complexes can fully displace BCR/Ag complexes and occupy the central region (of the order of millions of time steps, much longer than the biologically relevant time scale of a few hours).

(ii) $P_{A(BA)}^{max} > P_{A(LI)}^{max}$

Here the situation is just the opposite. The final equilibrium configuration will consist of a central cluster of high-affinity BCR/Ag complexes surrounded by a ring-shaped cluster of LFA-1/ICAM-1 complexes. Though this is the stable immunological synapse pattern observed in experiments, the time to reach this final segregation pattern is observed to be very long in our simulations (~ $10^9$ time steps). Even a single-species system consisting of high affinity BCR and antigen molecules can take a long time before collecting at the center of the region of binding. This is exacerbated by the addition of LFA-1 and ICAM-1 molecules, as the LFA-1/ICAM-1 complexes are trapped inside the high-affinity BCR/Ag complex cluster. This type of slow relaxation has been typically observed in a variety of "glassy" systems, where very similar kinetic trapping of particles in finite regions ("cages") is responsible for such long time relaxation time to equilibrium. However, no such glassy behavior was observed during synaptic experiments (6,7). In our simulations, it is necessary to incorporate either a signaling-driven affinity shift of LFA-1 or stiffer and/or shorter BCR/Ag bonds (relative to LFA-1/ICAM-1 bonds) in order to reach the equilibrium configuration within a biologically relevant time frame.

Figure S.1 Time evolution of a synapse pattern produced in Figure 4B, with BCR affinity $P^{max}_{A(BA)}=100$ ($K_A \approx 10^6$ M$^{-1}$) and $P^{max}_{A(LI)}=1000$ ($K_A \approx 10^7$ M$^{-1}$). The faster dissociation rate of the BCR/Ag complexes ensures that they have more or less formed a tight cluster by $t=10$ s, while the LFA-1/ICAM-1 complexes are still arranged in a ring-like structure after $t=100$ s. $N_{B/A}$ and $N_{L/I}$ refer to the number of BCR/Ag and LFA-1/ICAM-1 complexes formed, respectively

Figure S.2 Time evolution of the inverted synapse pattern produced in Figure 4D, with BCR affinity $P^{max}_{A(BA)}=10^4$ ($K_A \approx 10^8$ M$^{-1}$) and $P^{max}_{A(LI)}=1000$ ($K_A \approx 10^7$ M$^{-1}$). The faster dissociation rate of the LFA-1/ICAM-1 complexes ensures that they have more or less formed a cluster by $t=100$ s, while the BCR/Ag complexes remain in a ring-like structure even at $t=1000$ s, resulting in an inverted synapse pattern.

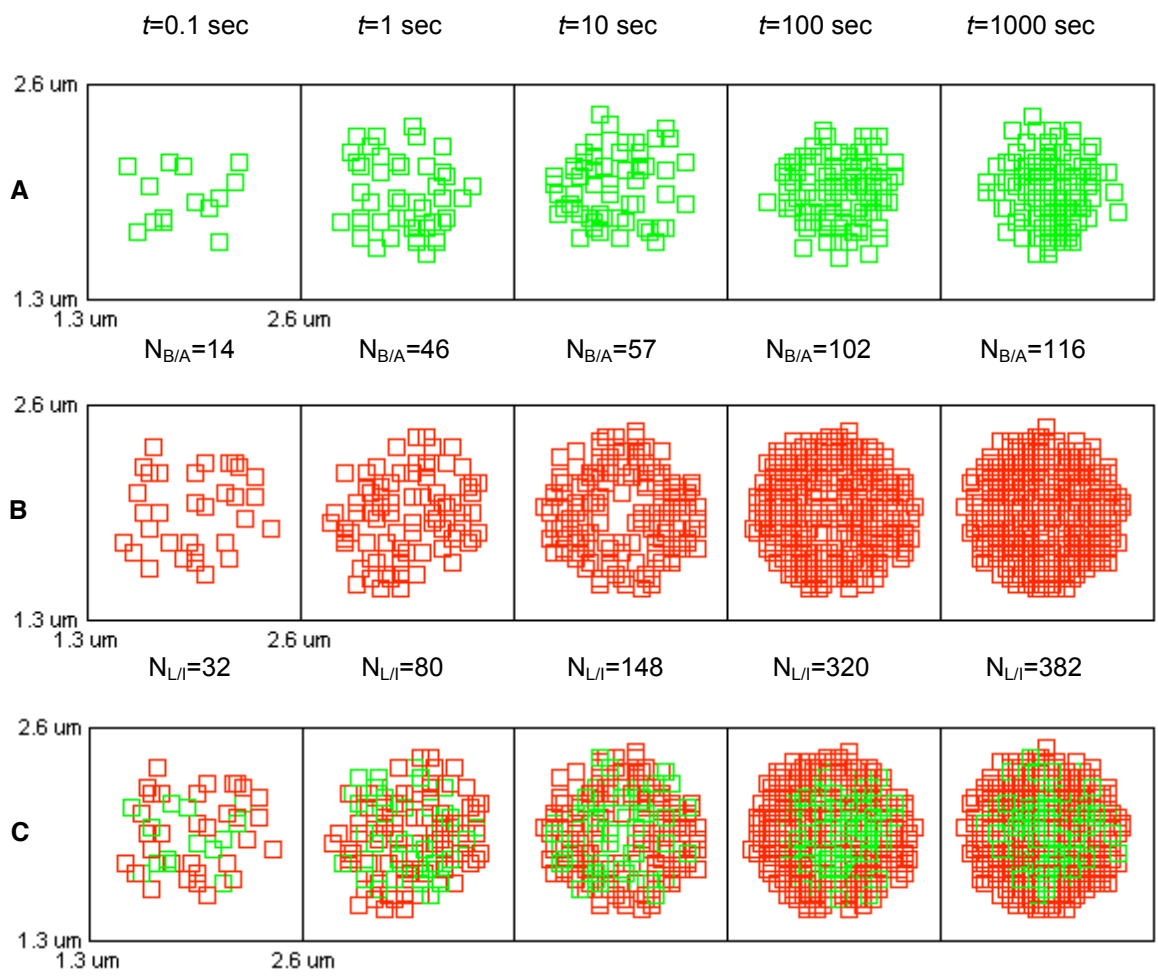

Figure S.1

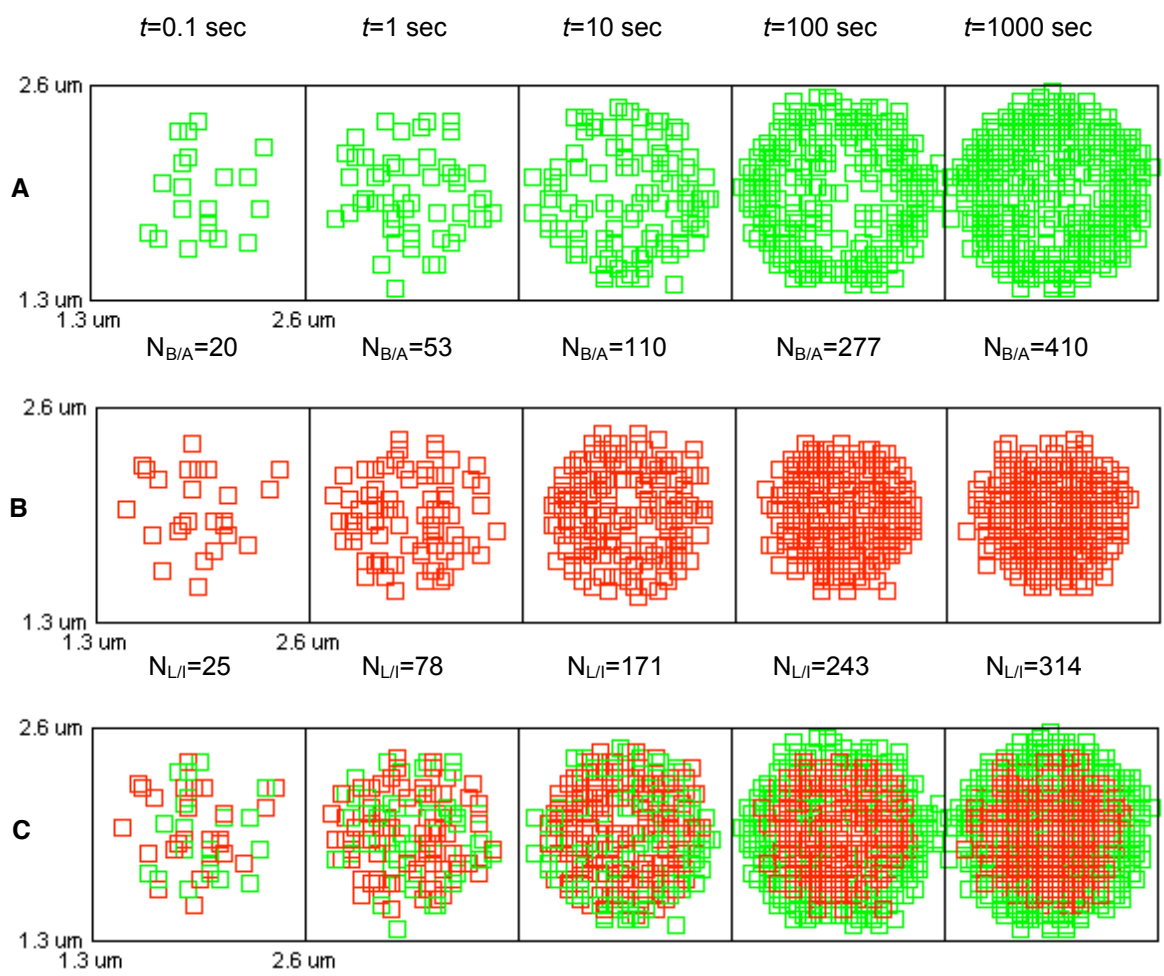

Figure S.2